\documentclass[copyright,creativecommons]{eptcs}
\usepackage{tikz}
\usetikzlibrary{shapes, positioning, decorations.pathreplacing, decorations.text, arrows, automata, fit, calc}
\usepackage{listings}
\usepackage{booktabs}
\usepackage{amsmath}
\usepackage{gensymb}
\usepackage{pgfplots}
\usepackage{color}
\pgfplotsset{compat=1.3}
\usepackage{filecontents}
\usepackage{array}
\newcolumntype{C}[1]{>{\centering\let\newline\\\arraybackslash\hspace{0pt}}m{#1}}

\definecolor{gray}{RGB}{160,160,160}
\definecolor{lightgray}{RGB}{224,224,224}
\definecolor{green}{RGB}{0,204,0}

\definecolor{darkblue}{rgb}{0.0,0.0,0.6}
\definecolor{cyan}{rgb}{0.0,0.6,0.6}

\tikzstyle{line} = [draw, -latex']
\tikzset{
    connector/.style={
     -latex,
     font=\scriptsize
    },
    rectangle connector/.style={
        connector,
        to path={(\tikztostart) -- ++(#1,0pt) \tikztonodes |- (\tikztotarget) },
        pos=0.5
    },
    rectangle connector/.default=-2cm,
    straight connector/.style={
        connector,
        to path=--(\tikztotarget) \tikztonodes
    }
}

\providecommand{\event}{V2CPS 2016} % Name of the event you are submitting to
\usepackage{breakurl}             % Not needed if you use pdflatex only.

\title{Towards the Verification of Safety-critical Autonomous Systems in Dynamic Environments}
\author{Adina Aniculaesei
\institute{TU Clausthal\\ 38678 Clausthal-Zellerfeld, Germany}
\email{adina.aniculaesei@tu-clausthal.de}
\and
Daniel Arnsberger 
\institute{TU Clausthal\\
38678 Clausthal-Zellerfeld, Germany}
\email{daniel.arnsberger@tu-clausthal.de}
\and
Falk Howar
\institute{TU Clausthal\\
38678 Clausthal-Zellerfeld, Germany}
\email{falk.howar@tu-clausthal.de}
\and
Andreas Rausch
\institute{TU Clausthal\\
38678 Clausthal-Zellerfeld, Germany} 
\email{andreas.rausch@tu-clausthal.de}
}
\def\titlerunning{Verification of Safety-critical Autonomous Systems}
\def\authorrunning{Aniculaesei, Arnsberger, Howar \& Rausch}
\lstdefinelanguage{XML}
{
  morestring=[b]",
  morestring=[s]{>}{<},
  morecomment=[s]{<?}{?>},
  stringstyle=\color{black},
  identifierstyle=\color{darkblue},
  keywordstyle=\color{cyan},
  morekeywords={xmlns,version,type}% list your attributes here
}

\lstdefinelanguage{UPPAAL}
{ % syntax highlight via font
	basicstyle=\small\sffamily, % small sans−serif font (like verdana)
	keywords={after update,assign,before update,break,case,const,continue,
		default,else,enum,for,guard,if,meta,process,progress,return,select,
		state,sync,switch,trans,system,while},
	keywords={[2]broadcast,bool,clock,chan,commit,init,int,scalar,struct,
		typedef,urgent,void}, 
	keywordstyle={[2]\bfseries},
	keywordstyle={[2]\color{black!50!green}},
	keywords={[3]false,true}, %otherkeywords={[3]−>},
	keywordstyle={[3]\color{magenta}},
	morekeywords={[3]−>}, 
	keywordstyle={[3]\bfseries},
	comment=[l]{//}, 
	morecomment=[s]{/∗}{∗/}, % single and multi−line
	commentstyle={\color{black!50!red}\itshape}, % appear in italic
	tabsize=4, % tab treatment (going to be fixed in Uppaal)
	captionpos=b, % put captions at the bottom
	escapechar=@ % write LaTeX comments escaped by @ symbol
}  % allows for indexgeneration

\begin{document}
\maketitle
%
% ------------------ Abstract -----------------------
%
\begin{abstract}
There is an increasing necessity to deploy autonomous systems in highly heterogeneous, dynamic environments, e.g. service robots in hospitals or autonomous cars on highways. Due to the uncertainty in these environments, the verification results obtained with respect to the system and environment models at design-time might not be transferable to the system behavior at run time. For autonomous systems operating in dynamic environments, safety of motion and collision avoidance are critical requirements. With regard to these requirements, Ma{\v{c}}ek et al. \cite{MVF+08} define the \emph{passive safety} property, which requires that no collision can occur while the autonomous system is moving. To verify this property, we adopt a two phase process which combines static verification methods, used at design time, with dynamic ones, used at run time. In the design phase, we exploit UPPAAL to formalize the autonomous system and its environment as timed automata and the safety property as TCTL formula and to verify the correctness of these models with respect to this property. For the runtime phase, we build a monitor to check whether the assumptions made at design time are also correct at run time. If the current system observations of the environment do not correspond to the initial system assumptions, the monitor sends feedback to the system and the system enters a passive safe state.
\end{abstract}

%
% ------------------ Introduction --------------------
%
\section{Introduction}\label{sec:introduction}
These days, autonomous systems are deployed mostly in known environments, e.g. industrial robot systems in production plants \cite{WBB+11}. If accurate models for the autonomous system and its environment can be obtained at design time, formal methods can offer strong guarantees for the system behavior with respect to specific correctness properties. At runtime, if the behavior of the system and its environment fit their respective models, then it is guaranteed that the system behavior satisfies the correctness properties formulated with respect to the system model at design time. However, there is an increasing necessity to deploy autonomous systems in highly heterogeneous, dynamic environments, e.g. service robots in hospitals or autonomous cars on the highways. Due the inherent uncertainty in these environments, the verification results obtained with respect to the design-time models might not be transferable to the system behavior at runtime.

Autonomous systems, such as mobile robots or autonomous vehicles belong to the spectrum of safety-critical applications. For this kind of systems, motion safety and collision avoidance are critical requirements. With regard to these requirements, the paper in \cite{MVF+08} introduces the notions of \emph{passive safety} and \emph{passive friendly safety}. The former ensures that the autonomous system does not actively collide with obstacles in its environment. In addition to this, the latter regards the system's environment as friendly and requires that the autonomous system maintains enough maneuvering room for the obstacles to avoid a collision as well. 

In this paper we present a novel concept for the verification of safety-critical autonomous systems in dynamic environments, with focus on the passive safety property. We establish the following premises for our verification problem. Firstly, the system's environment is dynamic and heterogeneous. Due to the dynamics of the environment, there is an infinite number of unforeseen situations, which cannot be modeled and verified at design time. Secondly, the autonomous system under verification is equipped with sensors, through which it can detect in real time the changes occurred in the environment. Furthermore, the system starts to run with predefined assumptions about its environment. The system safety property is verified at design time against the modeled system's behavior and the system's assumptions. %In this paper we present a concept for the verification of autonomous systems in dynamic environments. Using this concept, we can ensure that the system's safety property is satisfied for a larger subset of the runtime environment than the one covered by the system's assumptions about its environment.

Our concept presents a two-phase process for our verification problem. In the design phase, we make use of the UPPAAL model checker \cite{BDL04} to formalize the system and environment models as timed automata and the system's safety property as a TCTL formula. After we verify the system model against its safety property, we use the system and the environment model to build a monitor which checks whether the system assumptions made at design time are also correct at runtime. Our case study and concept evaluation are built on the scenario of a mobile service robot driving towards a given goal in a simulation environment.

\subsection{Paper Structure}
The structure of this paper is as follows: Section \ref{sec:relatedwork} contains an overview on previous work; Section \ref{sec:concept} presents our concept for the verification of autonomous systems in dynamic environments. Section \ref{sec:scenario} introduces the example scenario and Section \ref{sec:casestudy} discusses the case study. The results of our evaluation can be found in Section \ref{sec:evaluation}. We draw conclusions and discuss future work in Section \ref{sec:conclusion}.

%
% ------------------ Related Work --------------------
%
\section{Related Work}\label{sec:relatedwork}
Prior approaches \cite{AKS07,BFS12,MGP13,PYR+15} focus on modeling and verification techniques to ensure collision safety for autonomous systems in dynamic or unknown environments. Other works \cite{MP14} present a more including approach, which reflects on the whole spectrum of cyber-physical systems. We consider these papers in turn and discuss the difference to our concept. Bouraine et al. \cite{BFS12} address the problem of passive motion safety of a mobile robot with limited field-of-view deployed in an unknown environment. The authors introduce the notion of braking inevitable collision states which lead to collision regardless of the robot's trajectory. The navigation scheme presented in the paper avoids these states in order to achieve collision safety in environments with moving obstacles.

Mitsch et al. \cite{MGP13} use theorem proving techniques to verify the dynamic window algorithm for autonomous robotic ground vehicles against the passive safety and passive friendly safety properties. Both properties are verified with respect to an environment which contains stationary as well as moving obstacles. The paper makes use of the differential dynamic logic \cite{Pla08} as a a modeling formalism for the hybrid models which describe the continuous physical motion of the robot as well as its discrete control choices.

The paper in \cite{MP14} presents the ModelPlex approach, which combines offline verification of CPS models with runtime validation in order to provide correctness guarantees for system executions at runtime. The method uses theorem proving with sound proof rules to synthesize three runtime monitors, i.e. model monitor, controller monitor and prediction monitor, from hybrid system models. The first monitor checks the system execution for deviations from the system model. The second monitor tests the current controller decisions of the system implementation for compliance with the system model, while the prediction monitor evaluates the worst-case safety impact of the current controller decisions with respect to the predictions of a bounded deviation plant model.  

Phan et al. \cite{PYR+15} present a runtime approach based on the Simplex architecture \cite{RDW+96}, to ensure collision-freedom for robots with limited field-of-view and limited sensing range in unknown environments, i.e. environments where the detailed shapes and the locations of the obstacles are not known in advance. The switching condition between the advanced controller and the baseline controller is computed using extensive geometry reasoning. The approach guarantees collision-freedom if the obstacles are stationary. However, the authors claim that the approach can be extended for environments containing moving objects to ensure passive safety, if a bound on the obstacle velocity is known.

Alami et al. \cite{AKS07} present an approach which computes the maximum velocity profile of a mobile robot moving over a planned trajectory in an environment with an arbitrary number of obstacles. Both robot dynamics and environment, as well as the constraints of the robot sensors are considered in the computation of the velocity profile. The planned profile indicates the maximum speeds that the robot may have along its path without colliding with any object which could intercept its future trajectory. The maximum possible velocity with which the mobile objects move in the environment is known in advance.

Kane et al. \cite{KCD+15} address the runtime verification of an ARV with black-box commercial-off-the-shelf components, which are not amenable to instrumentation. Instead, the authors propose an approach to passive monitoring of the target system, by generating high-level property constructs from the observed network state. The runtime monitoring algorithm developed in this paper incrementally takes as input a system state and a MTL formula and checks the state trace for violations. In order to give the system enough time to react to environment changes, the authors propose to reduce the formula as soon as possible using history summarizing structures and simplifications based on formula-rewriting. 

Similar to the presented papers, we work on the premise that the real environment is highly dynamic and heterogeneous. In addition to this, we consider that the autonomous system under verification has assumptions about the input it can receive from the environment, e.g. the maximum velocity of moving obstacles. However, the changes which may occur in the environment can invalidate these assumptions, and thus render the system behavior as non-conform with regard to its motion safety specification. None of the previous works regards explicitly the system assumptions about its environment and to what effect these assumptions can be exploited in order to provide correctness guarantees for the system behavior at runtime.

%
% ------------------ Concept --------------------
%
\section{Concept}\label{sec:concept}
In this section we present our concept for verification of safety-critical systems in dynamic environments. We developed our concept starting from the following premises:
\begin{itemize}
\item The environment is heterogeneous and the infinite number of possible situations cannot be modeled and verified at design time.
\item The system starts to run with predefined assumptions about its environment.
\item The system together with its assumptions has been verified at design time against its safety specification.
\item The system observes the changes in the environment in real time.
\end{itemize}

\begin{figure}[h!]
\centering
\includegraphics[width=\textwidth,height=0.8\textheight,keepaspectratio]{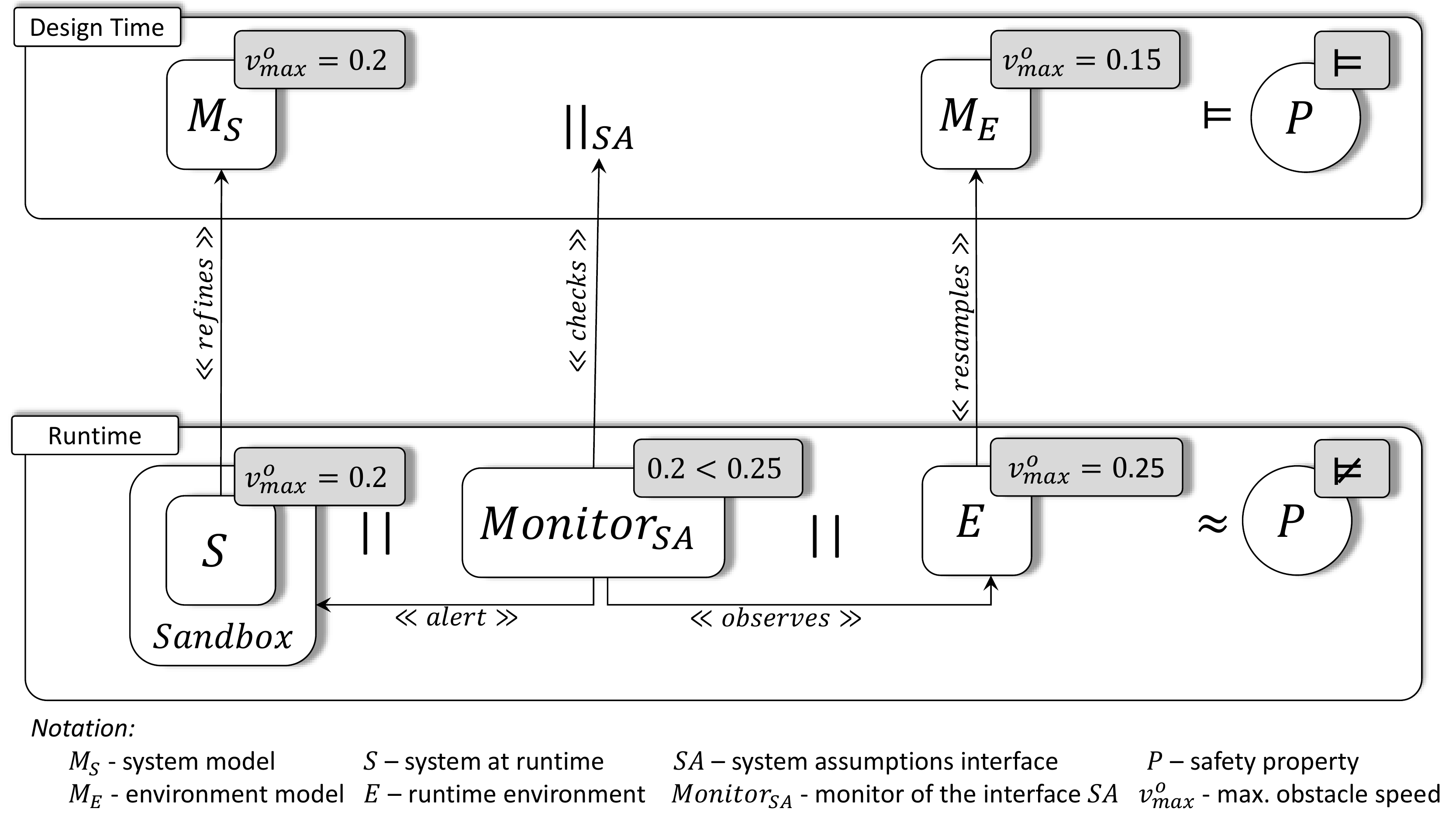}
\caption{Verification concept with runtime monitoring}
\label{fig:concept}
\end{figure}

We divide our concept in two parts: design time and runtime, as illustrated in Figure \ref{fig:concept}. 

At design time, $M_S$ describes the system behavior, while $M_E$ models the environment. The two models run in parallel and communicate with each other over an explicitly modeled interface $SA$. The passive safety property is formulated with regard to the system model $M_S$ and the environment model $M_E$ and is expressed as the property specification $P$. The environment model $M_E$ is constructed so that the behavior described in $M_S$ always satisfies $P$. To verify this, model checking is used as  verification method. 

We complement formal verification methods used at design time with runtime monitoring of the environment. In contrast to design time when all system executions can be inspected, only the current system execution can be verified at runtime against the safety property. The system $S$ refines the behavior described in its model $M_S$. During its runtime, the system $S$ operates in its environment $E$ and a monitor $Monitor_{SA}$ observes both in order to check whether the system assumptions made during design time are still valid at runtime. If the system assumptions are correct, the property specification $P$ is satisfied. Otherwise, the monitor gives feedback to the system and the system enters a safe state with respect to the passive safety property. According to the definition in \cite{MVF+08}, a system state \emph{s} is safe under the passive safety property if there exists at least one braking maneuver which in state \emph{s} and is collision-free for the duration of the braking time of the system.

In Figure \ref{fig:concept}, there are example values attached to each component involved in the verification problem. At design time, the system assumption about the environment is \textit{0.2} and the environment has a real upper bound of \textit{0.15}. Therefore, the property $P$ is indeed satisfied. At runtime, the environment has an upper bound of \textit{0.25} instead of \textit{0.15} while the system assumption remains unchanged. The monitor detects that the assumption is invalid and hence that $P$ is violated.
%\newpage

%
% ------------------ Scenario --------------------
%
\section{Scenario}\label{sec:scenario}
Our scenario takes place in a simulation environment in which a mobile service robot is commissioned to perform deliveries and reach a given destination. The environment features both stationary and moving objects.

\begin{figure}[h!]
\centering
\includegraphics[width=\textwidth,height=0.8\textheight,keepaspectratio]{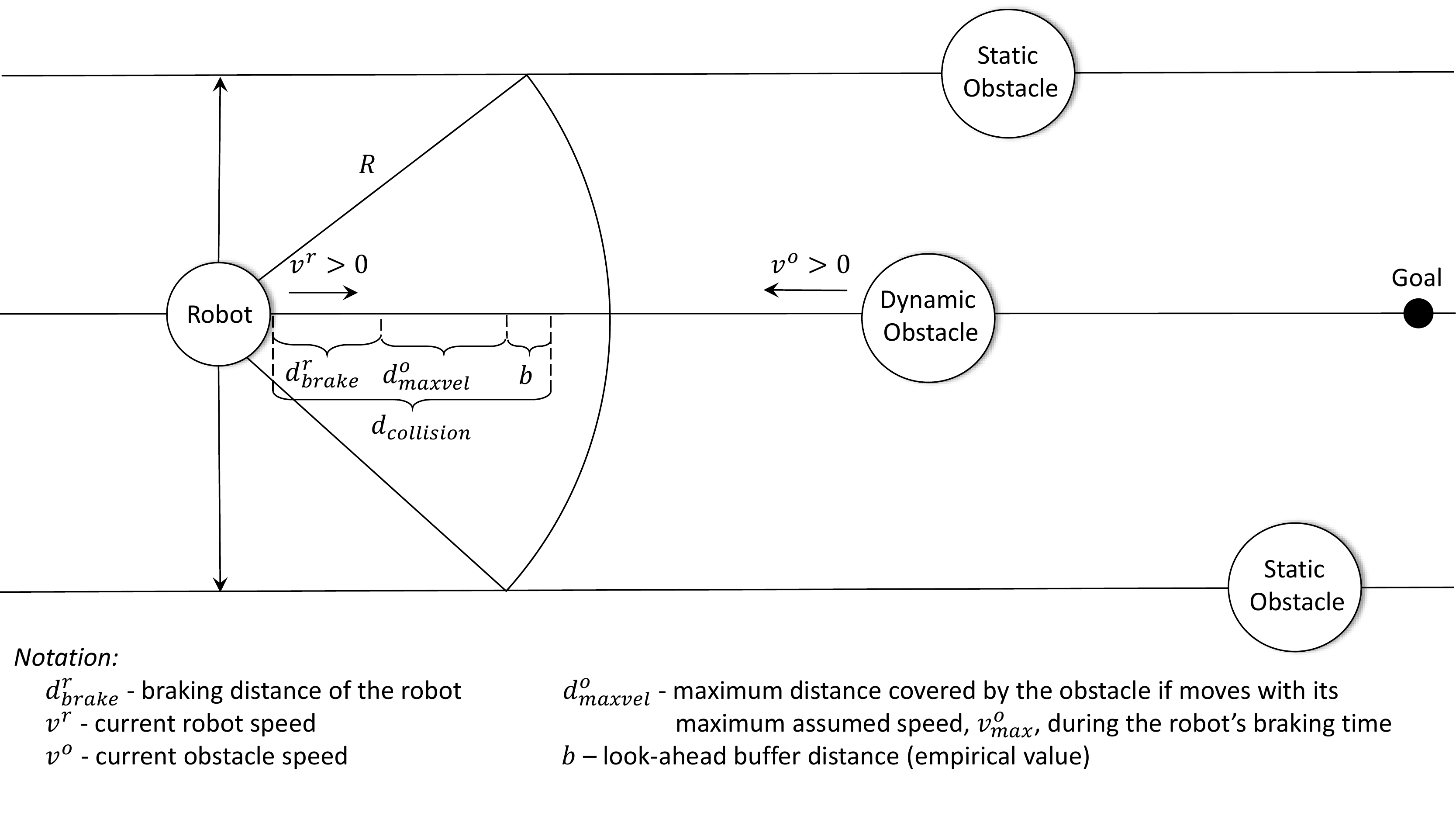}
\caption{Scenario and Physical Model.}
\label{physical-model}
\end{figure}

Figure \ref{physical-model} depicts an abstract view of the environment, in which the robot drives to its goal, while an obstacle moves on the same lane from the opposite direction. Static obstacles occupy the neighboring left and right lanes. The robot has partial knowledge of its surroundings due to its sensory limitations, i.e. its field of view spans up to range $R$. 

%\begin{figure}[!h]
%	\centering
%	\includegraphics[width=\textwidth]{./img/PhysicalModel}
%	\caption{Scenario: Physical Model.}
%	\label{fig:physical-model}
%\end{figure}

There are a few constraints we have imposed on the robot and on its environment, without affecting the generality of our concept. The robot and the dynamic obstacle move towards each other in a two-dimensional space. The robot has the ability to observe its environment and react to its changes. However, the robot reacts only to environment changes which occur in front of it. Even though it cannot rotate, the robot can change lanes by moving sideways.

As it starts to drive, the robot accelerates until it has reached its maximum velocity. Then, it continues to drive with this velocity until it reaches its destination or until it brakes due to collision danger. In order to detect a possible collision, the robot computes the distance $d_{collision}$ and compares it with the current distance between the robot and the moving obstacle. The distance $d_{collision}$ is calculated as follows: 
\begin{equation}
d_{collision} = d^r_{brake}+d^o_{maxvel}+b
\end{equation} where:
\begin{itemize}
\item $d^r_{brake} = v^r*t^r_{brake}$ is the braking distance of the robot based on its current velocity $v^r$,
\item $d^o_{maxvel} = v^o_{max}*t^r_{brake} $ is the distance covered by the obstacle moving with maximum velocity $v^o_{max}$ during the robot's braking time $t^r_{brake}$, and
\item $b$ is a look-ahead buffer distance, whose value is determined through experiments. This is necessary in order for the robot to stop before causing a collision with the obstacle.
\end{itemize}
When detecting a possible collision, the robot also checks for the possibility to change the lane rather than braking immediately.
Without reducing the generality of our concept, we assume that there is only one moving obstacle in the robot's environment. The obstacle's velocity is bounded by a maximum value, $v_o \in (0,v^o_{max}]$ and can change randomly in the interval boundaries given by $v^o_{max}$.

%
% ------------------ Case Study --------------------
%
\section{Case Study}\label{sec:casestudy}
In this section, we present the case study, on the basis of which we evaluate our concept. In the scenario introduced in Section \ref{sec:scenario}, we presented an informal specification for a mobile service robot, i.e. the robot must reach a specific point, while avoiding obstacles as well as possible by braking timely before a collision takes place. 

In order to check if the robot complies with its specification, we use model checking to formally verify the behavior of the robot against its specification. We use the UPPAAL model checker \cite{BDL04} to describe the robot's behavior and that of its environment, and to formalize the robot's specification. UPPAAL uses timed automata as its modeling language and Timed Computation Tree Logic (TCTL), a subset of Computation Tree Logic (CTL), as its specification language. 

\subsection{System and Environment Models}
\begin{figure}[h!]
\centering
\includegraphics[width=\textwidth,height=\textheight,keepaspectratio]{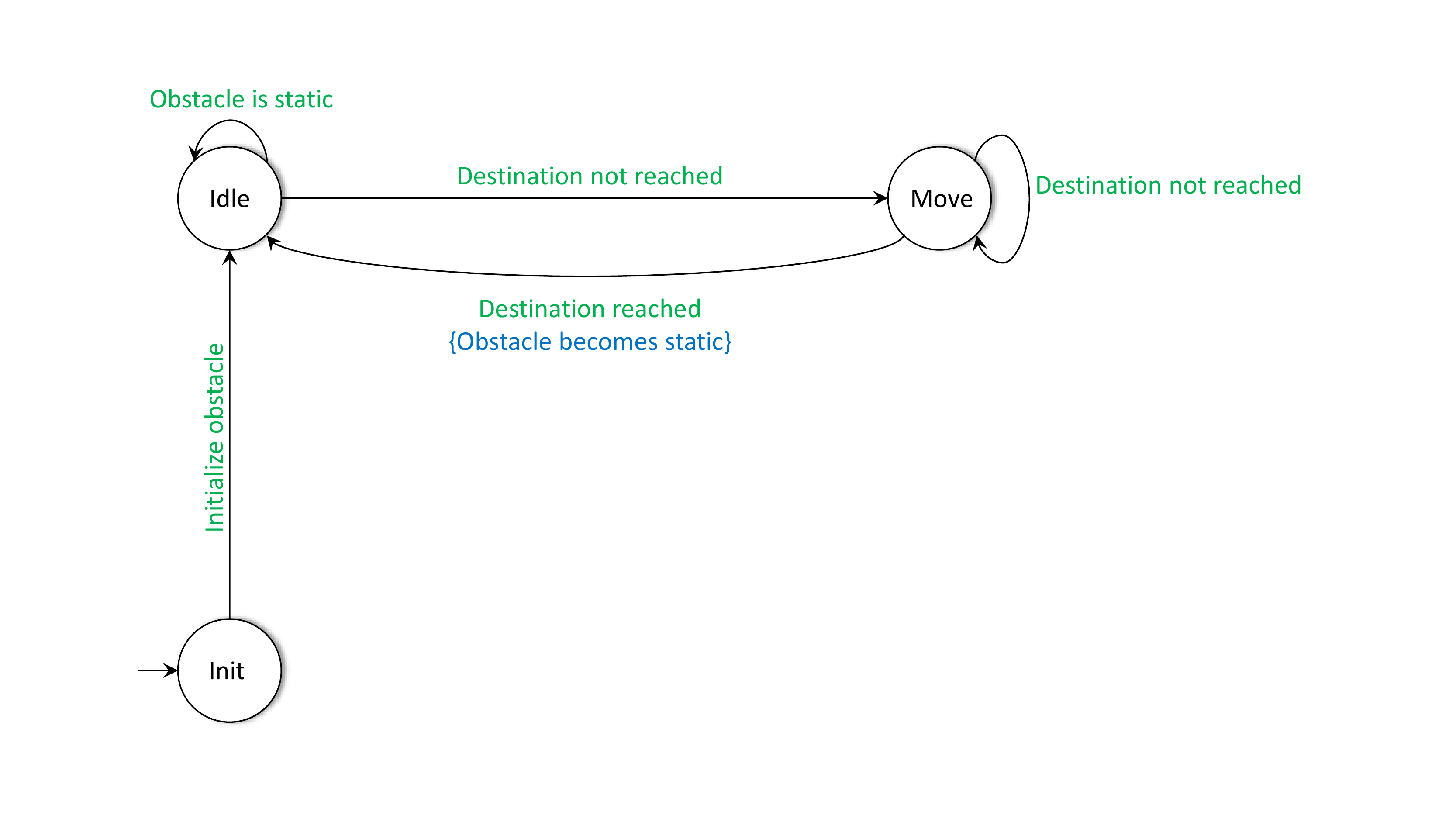}
\caption{Abstract obstacle automaton}
\label{fig:uppaal_obstacle}
\end{figure}

\begin{figure}[h!]
\centering
\includegraphics[width=\textwidth,height=\textheight,keepaspectratio]{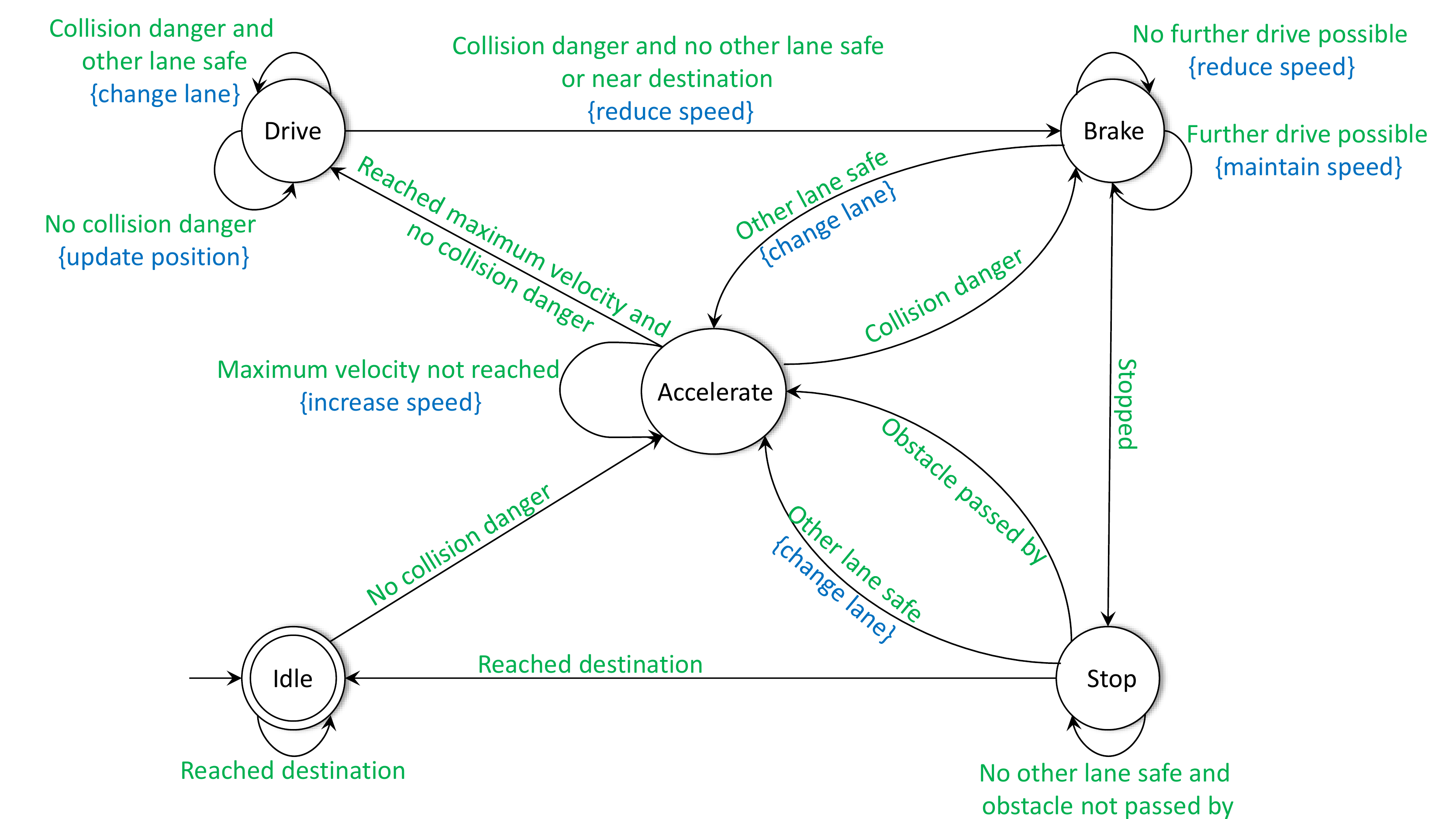} 
\caption{Abstract robot automaton}
\label{fig:uppaal_robot}
\end{figure}

In Figure \ref{fig:uppaal_obstacle} and Figure \ref{fig:uppaal_robot} two abstract automata are illustrated, which model the behavior of the robot and that of the obstacles in the environment. In order to simplify the models, we abstract from their characteristics as physical objects and consider the robot and the obstacles as discrete points in a two-dimensional space. The robot cannot drive through a moving obstacle, since this would correspond to a collision actively caused by the robot. However, in order to emulate the movement of the obstacle past the robot, we allow the discrete point representing the obstacle to move through the point which illustrates the robot.

The template of the obstacle automaton is instantiated accordingly in order to account for two types of obstacles in the environment: static and dynamic obstacles. The automaton has three locations, the initial location \textit{Init} and the locations \textit{Idle} and \textit{Move}. On the first transition from \textit{Init} to \textit{Idle}, the obstacle's parameters, i.e. identification number, initial position, whether it is static or not, are stored in a global obstacle array. In the \textit{Idle} state, an obstacle can be either static or it can begin to move towards the robot in a continuous motion. The obstacle velocity has an upper bound and the obstacle automaton chooses arbitrarily its velocity from the interval $(0,v^o_{max}]$, each time the obstacle executes a move. Based on the current velocity, the new obstacle position is computed and updated. The obstacle automaton is modeled so that a moving obstacle has a destination, and upon reaching it, the obstacle becomes static, i.e. the automaton enters the location \textit{Idle}. This modeled behavior ensures that the obstacle eventually stops moving.

The robot automaton is displayed in Figure \ref{fig:uppaal_robot} and it has the following locations: \textit{Idle}, \textit{Accelerate}, \textit{Drive}, \textit{Brake}, \textit{Stop}. The robot is stationary before starting its drive and when it has reached its destination. Therefore, the state \textit{Idle} is both initial and final state. As long as it has not reached its maximum speed, the robot will accelerate. The robot will continue to drive at full speed, provided no collision danger has been detected. In case of collision danger, the robot triggers the brake and checks whether it can drive further, albeit at reduced speed, or if it can change the lane. If neither lane change nor the further driving at reduced speed is possible, then the robot brakes until it comes to a full stop. 
%\begin{itemize}
%	\item \textbf{Idle} is the initial location of the robot automaton, because the robot is stationary at the beginning. When it reaches its destination, the robot stops, so this is also the final state.
%	\item \textbf{Accelerate} is the location in which the automaton remains as long as the robot has not reached its maximum speed.
%	\item \textbf{Drive} is the location in which the automaton remains as long as the robot can drive without any collision danger.
%	\item \textbf{Brake} is the location which the automaton enters if collision danger or wrong assumption is detected.
%	\item \textbf{Stop} is the location which the automaton enters when the robot stands still after braking. 
%\end{itemize}

The transitions from one location to another are guarded with different conditions (depicted in green), which express if the respective transition is enabled or not. Update statements (illustrated in blue) are used to change accordingly the values of various variables, e.g. robot speed or position. 

We elaborate only on one of the functions used in the transitions guards, the function \textit{collisionDanger} shown in Listing \ref{lst:collision-danger}. It calculates whether a possible collision is ahead or not. In order to perform its computation, the function considers the maximum obstacle velocity assumed by the robot.

\begin{lstlisting}[language={UPPAAL}, frameround=fftt, frame=shadowbox,rulesepcolor=\color{gray}, numbers=left,numberstyle=\tiny,stepnumber=1,numbersep=5pt, breaklines=true,
caption={Guard function which checks for collision danger},label=lst:collision-danger]
bool collisionDanger() {
	int i = 0;
	while (i < N) {
		if (robotLane == obstacles[i].lane) {		
			if (robotPosition <= obstacles[i].position && obstacles[i].position - robotPosition <= visualRadius) {
				if ((robotPosition + brakingDistance(vMax) + obstacleDrivingDistance(vMax)) >= obstacles[i].position - BUFFER && !obstacles[i].static) {
					return true;
				} 
				else if (robotPosition + brakingDistance(vMax+1) >= obstacles[i].position && obstacles[i].static) {
					return true;
				}
			}
		}
		i++;
	}
	return false;
}

\end{lstlisting}

The function checks for all obstacles in the environment if the obstacle is on the same lane as the robot and if a collision is possible. The computations for the detection of collision danger are performed if and only if the obstacle has entered the robot's visual range. Two other functions are used in this computation: \textit{brakingDistance} and \textit{obstacleDrivingDistance}. The first function computes the robot's braking distance for its maximum velocity, while the latter calculates the distance which the robot assumes the obstacle can still drive during its braking time. We further add a look-ahead buffer distance, in order to account for the fact that the robot gets the current informations about the obstacle one time step after the update is sent. Therefore, the collision distance between the robot and the obstacle is an upper bound. We also distinguish between dynamic and static obstacles, hence the two different computations. If there is a static obstacle in front of the robot, the computation is performed using only the robot's own speed.

\subsection{Safety Property}

The robot specification states that it must at all times comply with its safety property, namely never actively collide with an obstacle. This is expressed in Equation \ref{eq:safety-constraint}, which translates to the robot having no speed at the moment when an obstacle and the robot occupy consecutive positions on the same lane.
\begin{equation}\label{eq:safety-constraint}
\begin{aligned}
A[] \ forall \ (i:int[0,N-1]) & \ R.y \ == \ obstacles[i].y \\ & and \ obstacles[i].x \ > \ R.x \\ & and \ (obstacles[i].x \ - \ R.x \ <= \ 1) \\ & imply \ R.v \ == \ 0
\end{aligned}
\end{equation}
The satisfaction of this specification depends on the robot's assumption about the maximum obstacle speed. If the assumption is correct or greater than the real value, then the specification is satisfied and the robot stops in time. Otherwise, the robot brakes too late and a collision cannot be avoided.

\subsection{From UPPAAL Models to Implementation}

In order to perform the runtime analysis, the behavior described in the robot and obstacle models had to be transferred in executable source code. The transformation from UPPAAL models to executable source code was done using automata-based programming. Without reducing the generality of our concept, we implemented the robot without the functionality of changing lanes. Thus, we focused only on the timely braking of the robot, namely on the conditions in which this takes place, i.e. robot parameters or environment inputs.

\begin{table}[h!]
\centering
\begin{tabular}{|C{0.4\textwidth}|C{0.5\textwidth}|}
\toprule[0.2em]
\multicolumn{1}{|c|}{\textbf{UPPAAL}} & \textbf{Implementation}\\
\midrule
%\begin{lstlisting}[language=XML, basicstyle=\footnotesize]
%<location id="id0" x="561" y="-977">
	%<name x="510" y="-1003">Brake</name>
%</location>
%<location id="id1" x="187" y="-705">
	%<name x="93" y="-722">Accelerate</name>
%</location>
%<location id="id2" x="561" y="-433">
	%<name x="578" y="-458">Stop</name>
%</location>
%<location id="id3" x="-187" y="-977">
	%<name x="-170" y="-1003">Drive</name>
%</location>
%<location id="id4" x="-187" y="-433">
	%<name x="-221" y="-458">Idle</name>
%</location>
%\end{lstlisting}&
%\begin{figure}[h!]
\centering
%\begin{tikzpicture}[->,>=stealth', shorten >=1pt, auto, node distance=2cm, semithick]
%	\tikzstyle{every state}=[fill=lightgray, draw=none, text=black]
%	
%	\node[state]          (Accelerate) {\small $Accelerate$};
%	\node[state]          (Idle) [below left of=Accelerate] {\small $Idle$};
%	\node[state]          (Drive) [above left of=Accelerate] {\small $Drive$};
%	\node[state]          (Brake) [above right of=Accelerate] {\small $Brake$};
%	\node[state]          (Stop) [below right of=Accelerate] {\small $Stop$};
%\end{tikzpicture}
\includegraphics[width=0.4\columnwidth,keepaspectratio]{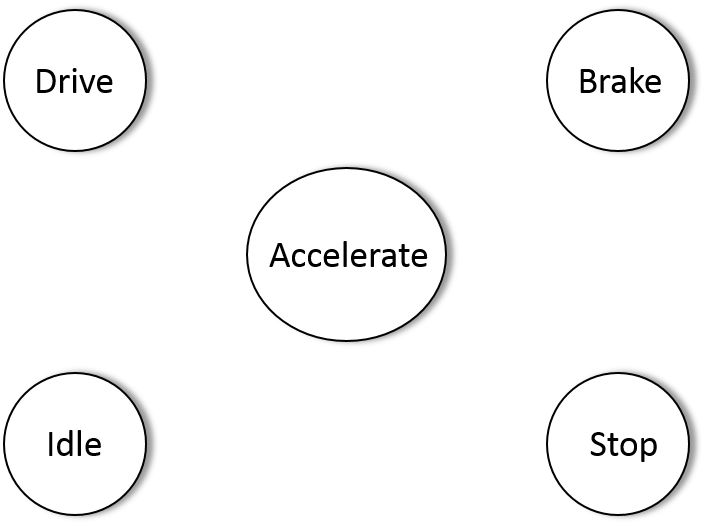}
%\end{figure}
&
\begin{lstlisting}[language=Python, basicstyle=\normalsize]
class RobotStates(Enum):
	Idle = 0
	Accelerate = 1
	Drive = 2
	Brake = 3
	Stop = 4
\end{lstlisting} \\ \hline
%\begin{lstlisting}[language=XML, basicstyle=\footnotesize, breaklines=true]
%<transition>
	%<source ref="id3"/>
	%<target ref="id0"/>
	%<label kind="guard" x="-42" y="-1028">(collisionDanger() && otherLaneSafe() == y)|| nearDestination() || wrongAssumption()</label>
	%<label kind="synchronisation" x="136" y="-1053">r?</label>
	%<label kind="assignment" x="85" y="-1011">updateLastSeen(), v = v-1, x = x+v</label>
%</transition>
%\end{lstlisting}&
\centering
%\begin{tikzpicture}[->,>=stealth', shorten >=1pt, auto, node distance=6.5cm, semithick]
%	\tikzstyle{every state}=[fill=lightgray, draw=none, text=black]
%	
%	\node[state]          (Drive) {\small $Drive$};
%	\node[state]          (Brake) [right of=Drive] {\small $Brake$};
%	\path (Drive) edge node[align=center, green, pos=0.5, sloped, above] {\footnotesize Collision danger and no other lane safe \\ \footnotesize or wrong assumption or \footnotesize near destination \\ \footnotesize \textcolor{blue}{(reduce speed)}} (Brake);
%	
%\end{tikzpicture}
\includegraphics[width=0.4\columnwidth,keepaspectratio]{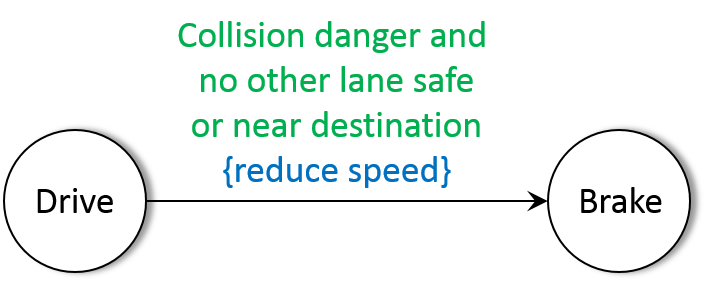}
%\end{figure}
&
\begin{lstlisting}[language=Python, basicstyle=\footnotesize, breaklines=true]
if (self.state == RobotStates.Drive and 
	 (self.collisionDanger() or 
	 self.nearDestination())):
		self.current_velocity_x -= self.velocity_increment
		self.state = RobotStates.Brake
\end{lstlisting}\\ \hline
\end{tabular}
\caption{Transformation from UPPAAL models to Python source code.}
\label{table:models-python}
\end{table} 

Table \ref{table:models-python} shows two examples of how the UPPAAL models are transformed into Python code. The first row shows how the five different robot states are declared. On the left side, the five states of the robot automaton are depicted graphically, while on the right side, the states are encoded with an enumeration structure. The second row exemplifies the transition step of the robot automaton through the transition between the states \textit{Drive} and \textit{Brake}. The functions \textit{collisionDanger} and \textit{wrongAssumption} implement the functionality specified in the UPPAAL model. We consider only one lane at runtime, as we chose not to implement the functionality of changing lanes in the robot. The same kind of transformation was performed for all the other transitions of the robot automaton as well as for the obstacle automaton.
%\newpage

%
% ------------------ Evaluation --------------------
%
\section{Evaluation and Discussion}\label{sec:evaluation}
In order to evaluate the approach presented in Section \ref{sec:concept}, we built the scenario in a simulation environment for robots and implemented the behavior which we have modeled in the case study. 

We built a test suite, in which we chose the robot's assumption about the maximum obstacle velocity to be always 0.2 m/s. We experimented with various maximum obstacle velocities as well as different radii of the robot's reaction area. When it observes a wrong assumption inside its reaction area, the robot begins to brake in order to enter a passive safe state. This area can be smaller than the visual range covered by the robot's sensors. As evaluation criterion for our concept we use the number of collisions which take place.

\begin{figure}[h!]
\centering
%\hspace*{2cm} 
\begin{tikzpicture}
%20, 26
\begin{axis}[view={150}{35},axis lines=left,mesh/cols=4, grid=major,
grid style={dotted}, xlabel={Max. Obstacle Velocity},
ylabel={Reaction Radius}, 
zlabel={Collisions},
ylabel style={xshift=-0.1cm, yshift=0cm, rotate=-45},
zlabel style={xshift=0.3cm},
xtick={0.2,0.25,0.3},
ytick={1.2,1.4,1.6},
ztick={0,2,4,6,8,10},
xmin=0.14,xmax=0.3,
zmin=0,zmax=10,
ymin=1,ymax=1.6,]
\addplot3[mark=none,surf,point meta=z, shader=flat, opacity=0.5] table[x index=0, y index=1,z index=2] {dataReact.dat};
%\addplot3[mark=none,surf, color=green, shader=flat] table[x index=0, y index=1,z index=2] {dataNoReact.dat};
\end{axis}
\end{tikzpicture}
\caption{Evaluation Results}
\label{fig:evalResults}
\end{figure}
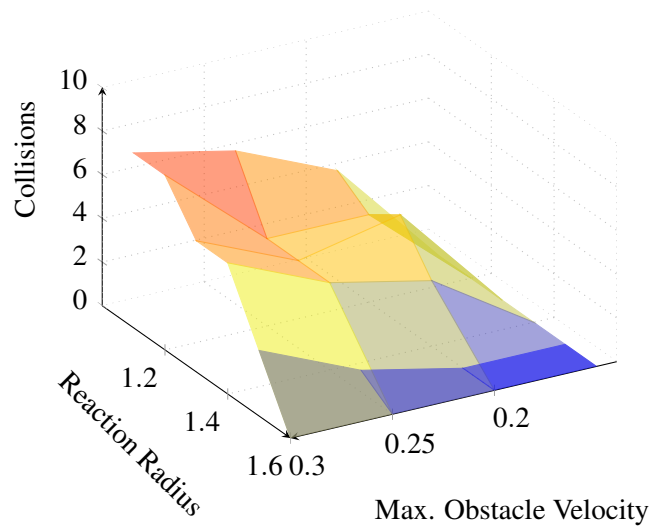

Figure \ref{fig:evalResults} shows the tests results. The blue region illustrates the tests which were already covered through model checking at design time. The maximum obstacle velocity did not exceed the robot's assumption and the reaction area was large enough for the robot to brake in time. The test cases in which the collisions took place are ordered in ascending order after the number of collisions and distributed in color-coded regions, from yellow to red, accordingly. When the reaction radius was chosen too small, the robot could not avoid a collision, even if its assumption was not exceeded by the obstacle. Furthermore, collisions took place also when the robot's reaction radius was chosen large enough, but the maximum obstacle speed was too high. Nevertheless, we were able to reduce the number of collisions by choosing an appropriate reaction radius for the different maximum obstacle velocities.

%
% ------------------ Conclusions and Future Work --------------------
%
\section{Conclusions and Future Work}\label{sec:conclusion}
This paper proposed the use of runtime monitoring to complement formal methods in the verification of safety-critical autonomous systems in heterogeneous environments. We developed a two-phase process to verify the behaviour of a mobile service robot in a simulation environment with respect to the passive safety property. In the design phase, we formalized the system and the environment models as timed automata in UPPAAL. We verified these models along with the system's assumptions about the environment against the system's safety property expressed as a TCTL formula. At runtime, we built a runtime monitor to check whether the system assumptions made at design time are still correct at runtime.

%Our evaluation showed that for a small enough reaction area, the robot could not avoid a collision even if the assumption made at design time were correct at runtime. Moreover, collisions also took place when the robot's reaction range was large enough, due to the major deviation between the system assumptions and the system observations. Still we were able to reduce the number of collisions by choosing a reaction area in accordance with the respective maximum obstacle velocity.

Our evaluation results show that we were able to ensure the satisfaction of the system's safety property for a larger subset of the runtime environment than the one covered by the assumptions which the system has about this environment. The size of this subset depends not only on the maximum obstacle velocity, but also on the reaction area of the robot. 

In future work we want to explore possible dependencies between system's assumptions and the effects one or a subset of incorrect assumptions may have on the runtime system's behavior. In this work, the implementation of the functionality described in the system model was done manually. Automated model transformation methods can be applied to extract the system implementation more easily from the system model. However, this transformation must in turn be verified to ensure it is performed correctly. In this respect, we want to identify a suitable verification method for model transformation (see \cite{CS13} for an extensive survey) and apply it to our case study. Furthermore, we intend to expand to several dynamic obstacles in the runtime environment and include sideway motion in an arbitrary manner for the dynamic obstacles, e.g. obstacles crossing the path of the autonomous system. Another aspect we want to consider is refined kinematic capabilities for the autonomous system, e.g. rotating, driving backwards or being outrun by moving obstacles.  

\nocite{*}
\bibliographystyle{eptcs}
\bibliography{generic}
\end{document}